\documentclass[pra,twocolumn,aps,amssymb,showpacs,superscriptaddress]{revtex4-1}

\usepackage{epsfig}
\usepackage{graphicx}
\usepackage{amsmath}
\usepackage{amssymb}
\usepackage{amsmath,amssymb}
\usepackage{enumerate}
\usepackage{hyperref}
\usepackage[normalem]{ulem}
\usepackage{cancel}
\usepackage{comment}

\usepackage{color}
\definecolor{rosso}{rgb}{1,0,0}
\definecolor{verde}{rgb}{0,1,0}
\definecolor{blue}{rgb}{0,0,1}
\definecolor{verdescuro}{rgb}{0,0.5,0.5}
\definecolor{rossoscuro}{rgb}{0.7,0.3,0}
\definecolor{bluscuro}{rgb}{0.3,0,0.7}
\definecolor{magenta}{rgb}{1,0,1}

\begin{document}

\title{Beyond-mean-field description of a trapped unitary Fermi gas \\ with mass and population imbalance}

\author{M. Pini}
\email{michele.pini@unicam.it}
\affiliation{School of Science and Technology, Physics Division, Universit\`{a} di Camerino, 62032 Camerino (MC), Italy}
\author{P. Pieri}
\affiliation{Dipartimento di Fisica e Astronomia, Universit\`a di Bologna, I-40127 Bologna (BO), Italy}
\affiliation{INFN, Sezione di Bologna, I-40127 Bologna (BO), Italy}
\author{R. Grimm}
\affiliation{Institut f\"ur Experimentalphysik, Universit\"at Innsbruck, 6020 Innsbruck, Austria}
\affiliation{Institut f\"ur Quantenoptik und Quanteninformation (IQOQI), \"Osterreichische Akademie der Wissenschaften, 6020 Innsbruck, Austria}
\author{G. Calvanese Strinati}
\email{giancarlo.strinati@unicam.it}
\affiliation{School of Science and Technology, Physics Division, Universit\`{a} di Camerino, 62032 Camerino (MC), Italy}
\affiliation{INFN, Sezione di Perugia, 06123 Perugia (PG), Italy}
\affiliation{CNR-INO, Istituto Nazionale di Ottica, Sede di Firenze, 50125 (FI), Italy}


\begin{abstract}
A detailed description is given of the phase diagram for a two-component unitary Fermi gas with mass and population imbalance, for both homogeneous and trapped systems.  
This aims at providing quantitative benchmarks for the normal-to-superfluid phase transition of a mass-imbalanced Fermi gas in the temperature-polarization parameter space. 
A self-consistent \mbox{$t$-matrix} approach is adopted, which has already proven to accurately describe the thermodynamic properties of the mass and population balanced unitary Fermi gas. 
Our results provide a guideline for the ongoing experiments on heteronuclear Fermi mixtures. 
\end{abstract}

\maketitle

\section{Introduction} 
\label{sec:introduction}

Ultra-cold gases provide a unique platform to investigate the nature of  pairing in fermionic superfluids. 
Through the use of Fano-Feshbach resonances \cite{Chin-2010}, it is possible to tune the attractive interaction between two fermionic components of the gas, making the superfluid system to span the BCS-BEC crossover from a 
Bardeen-Cooper-Schrieffer (BCS) condensate of highly overlapping Cooper pairs to a Bose-Einstein condensate (BEC) of dilute tightly-bound molecules \cite{Physics-Reports-2018}.
 At the resonance, the scattering length of the inter-particle interaction diverges and the system enters the so-called unitary regime. 
 
Besides the tunability of the interaction, ultra-cold Fermi gases allow for a direct control of the population of the two fermionic components undergoing pairing.
In the original experimental observations on the BCS-BEC crossover, which were realized with ultra-cold gases of fermionic $^{40}$K \cite{Greiner-2003} and $^6$Li \cite{Jochim-2003a,Jochim-2003b,Zwierlein-2003}, pairing was achieved with equal populations of two different hyperfine states of the same atomic species.
Afterwards, the effect of the population imbalance of the two components was also investigated experimentally \cite{Zwierlein-2006,Partridge-2006,Shin-2008a}.
Moreover, different fermionic species are available to realize combinations of different nature.
The first attempts to realize a mass-imbalanced mixture were made with a $^{6}$Li-$^{40}$K mixture \cite{Taglieber-2008,Wille-2008,Voigt-2009}. 
In this case, the occurrence of enhanced inelastic collisions near the Fano-Feshbach resonances \cite{Naik-2011} made the study of equilibrium properties rather problematic. 
More recently, two additional mass-imbalanced mixtures were considered, made of $^{40}$K-$^{161}$Dy \cite{Ravensbergen-Grimm-2018, Ravensbergen-Grimm-2020} and $^{6}$Li-$^{53}$Cr \cite{Neri-Zaccanti-2020}. 
In particular, a broad Fano-Feshbach resonance with suppressed inelastic processes was found in the K-Dy mixture \cite{Ravensbergen-Grimm-2020}, making it a promising candidate for realizing a mass-imbalanced superfluid.

On the theoretical side, the first works addressing pairing in mass-imbalanced Fermi mixtures were based on a mean-field approach 
\cite{Liu-2003,Wu-2003,Bedaque-2003,Lin-2006,Wu-Pao-Yip-2006,Wu-Pao-Yip-2007,Parish-2007,Gubbels-2009,Baarsma-2010,Wang-2017},
which correctly describes the system in the BCS (weak-coupling) limit but allows only for a qualitative description in the unitary regime.
These works resulted in a very rich phase diagram, both at zero temperature 
\cite{Liu-2003,Wu-2003,Bedaque-2003,Lin-2006,Wu-Pao-Yip-2006,Wu-Pao-Yip-2007,Parish-2007} 
and at finite temperature \cite{Parish-2007,Gubbels-2009,Baarsma-2010,Wang-2017}.
In particular, at finite temperature the phase diagram is governed by either a tricritical point (where the normal-to-superfluid phase transition changes from second to first order) or a Lifshitz point (where the order parameter of the superfluid in the normal-to-superfluid phase transition changes from being homogeneous to a Fulde-Ferrel-Larkin-Ovchinnikov (FFLO) periodic structure \cite{Casalbuoni-Nardulli-2004}), depending on mass ratio and coupling \cite{Parish-2007,Gubbels-2009,Baarsma-2010,Wang-2017}.

A first attempt to produce a beyond-mean-field finite-temperature phase diagram for a mass-imbalanced unitary Fermi gas was done in Refs.~\cite{Gubbels-2009,Baarsma-2010}, by using a polaronic \emph{ansatz} for the fermionic self-energy fitted on the zero-temperature quantum Monte Carlo equation of state of Ref.~\cite{Gezerlis-2009}, together with a screening correction to the interaction. 
The beyond-mean-field nature of the phase transition close to zero temperature has also been investigated in Refs.~\cite{Zdybel-2018,Zdybel-2019,Zdybel-2020} by the functional-renormalization-group approach.

Quite recently, beyond-mean-field \mbox{$t$-matrix} approaches were also employed to obtain a mass-imbalanced phase diagram with equal populations, first with partial \cite{Hanai-JLTP-2013,Hanai-PRA-2013} 
and later with full self-consistency \cite{Hanai-JLTP-2014,Hanai-PRA-2014}.
In this context, it is important to remark that \mbox{$t$-matrix} approaches are \emph{ab initio} and do not make use of phenomenological parameters. 
In particular, the fully self-consistent $t$-matrix approach (also known as Luttinger-Ward approach) \cite{Haussmann-1993,Haussmann-1994,Haussmann-2008, PPS-2019} appears to be the most suitable candidate for a quantitative description of the phase diagram of the  mass-imbalanced unitary Fermi gas. 
This approach compares rather well with Monte Carlo and experimental data for thermodynamic quantities at unitarity in the mass-balanced case \cite{Haussmann-2008,PPS-2019,Ku-Zwierlein-2012,Jensen-2020,Pini-2020}.  
In addition, when extended to the mass-imbalanced case, it does not suffer from the problem of a vanishing critical temperature on the BCS side of unitarity, which occurs in other non-self-consistent or partially self-consistent approaches \cite{Hanai-JLTP-2014,Hanai-PRA-2014}.

In this article, we use the self-consistent \mbox{$t$-matrix} approach to investigate the phase diagram of a unitary Fermi gas with \emph{both} mass and population imbalance.
To this end, we consider both a homogeneous and a harmonically trapped system. 
The explicit aim is to provide the optimal parameters that allow for the observation of superfluidity in experiments with mass-imbalanced Fermi gases. 
In the homogeneous case, we consider both a mixture of $^{40}$K and $^{161}$Dy atoms (like in Refs.~\cite{Ravensbergen-Grimm-2018,Ravensbergen-Grimm-2020}) and a mixture of $^{6}$Li and $^{53}$Cr atoms (like in Ref.~\cite{Neri-Zaccanti-2020}).
In the trapped case, we consider the K-Dy mixture only. 
Our analysis focuses on a second-order normal-to-superfluid phase transition.
Accordingly, it does not consider the occurrence of either phase separation related to a first-order normal-to-superfluid phase transition or an FFLO phase.

Our main results concern the temperature-polarization phase diagrams calculated at unitarity within the self-consistent $t$-matrix approach, for 
(i) a homogeneous K-Dy mixture, (ii) a homogeneous Li-Cr mixture, (iii) a harmonically trapped K-Dy mixture with two different trap configurations, and (iv) a harmonically trapped mass-balanced Fermi gas for reference purposes.
In particular, for the trapped K-Dy mixture we also investigate the dependence of the phase diagram on the ratio of the trap frequencies for the two atomic species, and characterize the phase diagram by varying this ratio. 
Experimentally, this can be done by variation of the wavelength of the trapping light or by employing a two-color trap, which selectively traps the two components with lasers of different wavelengths. 
We further present the density profiles in the trap, both at the critical temperature and in the normal phase. 
Finally, we quantify the increase of intra-species three-body recombinations which is due to the contraction of the density profiles with respect to the non-interacting case, within the experimental conditions of Ref.~\cite{Ravensbergen-Grimm-2020}.

The paper is organized as follows. 
Section~\ref{sec:theo} recalls the basic equations of the self-consistent $t$-matrix approach. 
Sections~\ref{sec:homo} and \ref{sec:trap} report on the numerical results obtained for the homogeneous and harmonically trapped systems, respectively. 
Section~\ref{sec:conclusion} gives our conclusions.

\vspace{-0.4cm}
\section{Theoretical approach}
\label{sec:theo}

In this Section, we introduce the basic equations of the self-consistent $t$-matrix approach for a homogeneous Fermi gas with mass and density imbalance. In the following, we shall set $\hbar=1$. 

We consider a mass-imbalanced Fermi gas at temperature $T$, with the light (L) and heavy (H) components interacting with each other through an attractive contact interaction with associated scattering length $a_{F}$.
The single-particle Green's function for the $\sigma$ component ($\sigma=L,H$) can be expressed as \cite{PPS-2019}
\begin{equation}
G_\sigma(k) = \Big(G_{0,\sigma}(k)^{-1} -\Sigma_{\sigma}(k)\Big)^{-1},
\label{eq:G_k}
\end{equation}
where $G_{0,\sigma}(k)=[\mathbf{k}^2/(2m_\sigma)-\mu_\sigma - i \omega_n]^{-1}$ is the non-interacting counterpart, with $m_\sigma$ the mass and $\mu_\sigma$ the chemical potential of the $\sigma$ component. 
Within the self-consistent $t$-matrix approach, the self-energy $\Sigma_{\sigma}(k)$ in Eq.~(\ref{eq:G_k}) is given by \cite{PPS-2019}
\begin{equation}
\Sigma_\sigma(k) = - \int \!\! \frac{d\mathbf{Q}}{(2\pi)^3} \frac{1}{\beta} \sum_\nu \Gamma(Q) \, G_{\bar{\sigma}}(Q-k) \, ,
\label{eq:Sigma_k}
\end{equation}
where
\begin{equation}
\Gamma(Q) = - \bigg(\frac{m}{4\pi a_F} + R_{\mathrm{pp}}(Q) \bigg)^{-1}
\label{eq:Gamma_Q}  
\end{equation}
is the particle-particle propagator with $m=2 m_L m_H/(m_L+m_H)$  twice the value of the reduced mass for the two components, and
\begin{equation}
R_{\mathrm{pp}}(Q) = \!\! \int \!\!\!\frac{d\mathbf{k}}{(2\pi)^3} \Big( \frac{1}{\beta} \sum_n G_\sigma(k) G_{\bar{\sigma}} (Q-k) - \frac{m}{\mathbf{k}^2} \! \Big),
\label{eq:Rpp_Q}
\end{equation}
the renormalized particle-particle bubble.
In the above expressions, $\beta=1/(k_B T)$ is the inverse temperature ($k_{B}$ being the Boltzmann constant), 
the index  $\sigma=(L,H)$ and its complementary one $\bar{\sigma}=(H,L)$ distinguish the two atomic species,
$k=(\mathbf{k},\omega_n)$ and $Q=({\mathbf{Q},\Omega_\nu})$ are four-vectors, 
where $\omega_n= (2n+1) \pi  \beta $ ($n$ integer) and $\Omega_\nu= 2 \pi \nu \beta $ ($\nu$ integer) are fermionic and bosonic Matsubara frequencies, respectively. 
Equations (\ref{eq:G_k})-(\ref{eq:Rpp_Q}) have to be solved up to self-consistency.
The numerical procedure to achieve this self-consistency is described in detail in Ref.~\cite{PPS-2019}.

The critical temperature $T_{c}$ for the normal-to-superfluid second-order phase transition is then obtained by the Thouless criterion \cite{Thouless-1960}:
\begin{equation}
\big[ \Gamma(Q=0;T=T_c) \big]^{-1} = 0 \, .
\label{eq:Thouless}
\end{equation}
This equation needs be solved together with the density equations for both $\sigma$ components
\begin{equation}
n_\sigma=-\int \frac{d \mathbf{k}}{(2\pi)^3} \frac{1}{\beta} \sum_n e^{i \eta \omega_n} G_\sigma(k) \quad (\eta \to 0^+),
\end{equation}
which determine the chemical potentials $\mu_\sigma$ for the two species with given densities $n_\sigma$.

\vspace{1cm}
\section{Numerical results for \\ the homogeneous system} 
\label{sec:homo}

In this Section, we consider a mass-imbalanced Fermi gas embedded in a homogeneous environment, whose knowledge will be needed for treating the trapped case discussed in Sec.~\ref{sec:trap}.  

The interaction between the two components is conveniently described in terms of the dimensionless coupling $(k_F a_F)^{-1}$ (at unitarity of interest here, $(k_F a_F)^{-1}=0$),
where the effective Fermi wave vector 
\begin{equation}
k_F=(3\pi^2 n)^{1/3}
\end{equation}
is defined in terms of the total particle density $n=n_L+n_H$. 
The effective Fermi energy 
\begin{equation}
\label{eq:E_F_homo}
E_F = \frac{(3\pi^2 n)^{2/3}}{2m} 
\end{equation}
can be also associated with this wave vector, where $m$ stands again for twice the value of the reduced mass like in Eqs.~(\ref{eq:Gamma_Q}) and (\ref{eq:Rpp_Q}). 
By rescaling the energies in terms of $E_F$, the lengths in terms of $k_F^{-1}$, and so on, the phase diagram becomes universal, meaning that, at fixed coupling $(k_F a_F)^{-1}$ and polarization $p=(n_H-n_L)/(n_H+n_L)$, the critical temperature $T_c$ 
(in units of the effective Fermi temperature $T_F=E_F/k_B$) becomes a function of the mass ratio $r_m=m_L/m_H$ only. 
In particular, $r_m=40/161$ for the K-Dy mixture of Refs.~\cite{Ravensbergen-Grimm-2018,Ravensbergen-Grimm-2020} and $r_m=6/53$ for the Li-Cr mixture of Ref.~\cite{Neri-Zaccanti-2020}). 

\begin{figure}[t]
\begin{center}
\includegraphics[width=7.5cm,angle=0]{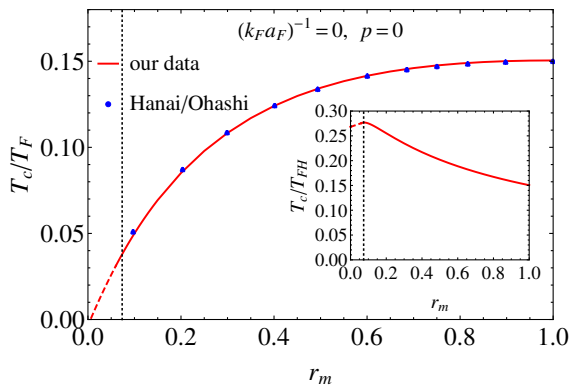}
\caption{(Color online)  The critical temperature $T_{c}$ (in units of the effective Fermi temperature $T_{F}$) obtained by  the self-consistent $t$-matrix approach (full line) is shown vs the mass ratio $r_m$ for a homogeneous density-balanced Fermi gas at unitarity,
                                      and compared with the theoretical values from Ref.~\cite{Hanai-JLTP-2014} (circles).
                                     The inset shows $T_{c}$ in units of the Fermi temperature $T_{FH}=(6\pi^2 n_H)^{2/3}/(2m_H k_B)$ of the heavy component.
                                     Both in the main panel and the inset, the dashed portions of the lines are extrapolations of the full lines toward $r_m=0$ and the dotted vertical lines correspond to the threshold for the Efimov instability at $r_m = 1/13.6$     
                                     \cite{Efimov-1970,Efimov-1973,Petrov-2003,Petrov-2005}.}
\label{Figure-1}
\end{center}
\end{figure} 

Figure~\ref{Figure-1} shows the ratio $T_c/T_F$ vs $r_m$ for a homogeneous Fermi gas at unitarity and with balanced densities ($p=0$). 
Our results agree with those obtained by a similar approach in Ref.~\cite{Hanai-JLTP-2014} (circles). 
The ratio $T_c/T_F$ is seen to decrease monotonically for decreasing $r_m$ and to extrapolate toward zero for $r_m=0$. 
However, at fixed $m_H$ this is just an effect of the chosen normalization in terms of $T_{F}$, since $T_{F} \to \infty$ for $r_m \to 0$. 
To remove this effect, in the inset of Fig.~\ref{Figure-1} the critical temperature is shown in units of the Fermi temperature $T_{FH}=(6\pi^2 n_H)^{2/3}/(2m_H k_B)$ of the heavy component. 
In this way, the ratio $T_c/T_{FH}$ is instead seen to increase for decreasing $r_m$, until it reaches a weak maximum for $r_m = 0.0763$ where $T_c/T_{FH}=0.277$, after which it decreases toward the (extrapolated) value $T_c/T_{FH}=0.268$ for $r_m\to 0$. 
It is interesting that the value of $r_m$ that corresponds to this maximum about coincides with the critical mass ratio $r_m = 1/13.6=0.0735$ (indicated by a vertical dotted line both in the main panel and the inset of Fig.~\ref{Figure-1}),
where the Efimov effect occurs and the system becomes unstable.
This instability occurs because light atoms can mediate an inverse square attractive potential between pairs of heavy atoms, which is strong enough to make the pairs to collapse \cite{Efimov-1970,Efimov-1973,Petrov-2003,Petrov-2005}. 

\begin{figure}[t]
\begin{center}
\includegraphics[width=7.5cm,angle=0]{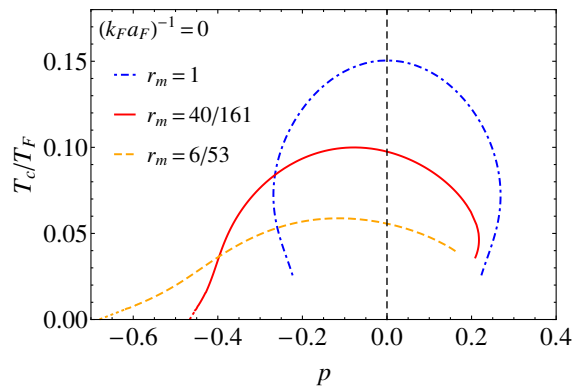}
\caption{(Color online) The critical temperature $T_c$ (in units of the effective Fermi temperature $T_F$) obtained by the self-consistent $t$-matrix approach is shown vs the polarization $p=(n_H-n_L)/(n_H+n_L)$ for the homogeneous system at unitarity.
                                     Several values of the mass ratio $r_m$ are considered, corresponding to the: 
                                    (i) mass-balanced Fermi gas ($r_m = 1$, dashed-dotted line), (ii) mass-imbalanced K-Dy mixture ($r_m=40/161$, full line), and (iii) mass-imbalanced Li-Cr mixture ($r_m=6/53$, dashed line).
                                     In the mass-imbalanced cases, the dotted portions of the lines represent linear extrapolations of the curves toward $T=0$ for negative $p$.}
\label{Figure-2}
\end{center}
\end{figure} 

The ratio $T_c/T_F$ can be also obtained as a function of the polarization $p$, by fixing the coupling $(k_F a_F)^{-1}$ and the mass ratio $r_m$.
This is shown in Fig.~\ref{Figure-2} for a homogeneous two-component Fermi system at unitarity, with
(i) equal masses (dashed-dotted line), (ii) the mass ratio $r_m=40/161$ corresponding to a K-Dy mixture (full line), 
and (iii) the mass ratio $r_m=6/53$ corresponding to a Li-Cr mixture (dashed line). 
We have verified that in the mass-balanced case our results agree with those obtained by a similar approach in Ref.~\cite{Frank-2018}.

Note from Fig.~\ref{Figure-2}  that in the mass-balanced case the maximum of $T_c/T_F$ occurs for equal densities ($p=0$). 
For mass-imbalanced mixtures, however, the maximum occurs instead at negative values of the polarization $p$, which corresponds to a majority of light atoms 
(this behavior is in line with what found within a mean-field approach in Refs.~\cite{Gubbels-2009,Baarsma-2010,Wang-2017,Footnote-1}).
In particular, from Fig.~\ref{Figure-2}  we obtain $p_\mathrm{max}=-0.0775$ for the K-Dy mixture and $p_\mathrm{max}=-0.114$ for the Li-Cr mixture. 
It should, however, be remarked that the value of $p_\mathrm{max}$ depends on the units used for normalizing $T_{c}$. 
When these units are in terms of the Fermi temperature $T_{F H}=(6\pi^2 n_H)^{2/3}/(2 m_H k_B)$ of the heavy component, we obtain $p_\mathrm{max}=-0.166$ for the K-Dy mixture and $p_\mathrm{max}=-0.204$ for the Li-Cr mixture;
otherwise, when the units are in terms of the Fermi temperature $T_{F L}=(6\pi^2 n_L)^{2/3}/(2 m_L k_B)$ of the light component, we obtain $p_\mathrm{max}=0.003$ for the K-Dy mixture and $p_\mathrm{max}=-0.036$ for the Li-Cr mixture.

Note further from Fig.~\ref{Figure-2}  that for the K-Dy mixture, on the $p>0$ side the $T_c$ curve decreases up to a point where it starts developing a reentrance, in a similar way to what happens for the mass-balanced case
(the reentrance points being $(p^+,T^+_c/T_F)=(0.2175,0.0462)$ for the K-Dy mixture and $(p^+,T^+_c/T_F)=(0.7180,0.0646)$ for the mass-balanced case). 
Usually, this reentrant behavior is associated with a region of phase separation in the phase diagram, where the normal-superfluid phase transition becomes of first-order and the second-order curve for $T_c$ is covered by the phase separation region \cite{Sarma-1963}. From mean-field calculations \cite{Gubbels-2009,Baarsma-2010,Wang-2017}, we expect in addition an FFLO phase to develop in this region of the phase diagram. 
For the Li-Cr mixture, on the other hand, we have not been able to find a reentrant region in the $T_c$ curve due to convergence problems in the numerical calculations. 
In this case, it might be possible that a reentrance point cannot be reached by our numerical algorithm   owing to an FFLO phase developing before the reentrance. 
Note finally from Fig.~\ref{Figure-2}  that on the $p<0$ side $T_c$ decreases monotonically with increasing $|p|$, until it reaches $T=0$ at the (extrapolated) critical values $p^{-} \simeq -0.47$ for the K-Dy mixture and $p^{-} \simeq -0.68$ for the Li-Cr mixture. 

As a final comment, we mention that at a \emph{qualitative\/} level  the unitary phase diagrams for the second-order phase transition shown in Fig.~\ref{Figure-2}, obtained by the self-consistent $t$-matrix approach, appear quite similar to those obtained in Refs.~\cite{Gubbels-2009,Baarsma-2010,Wang-2017} by a mean-field approach. 
However, at a \emph{quantitative\/} level significant differences occur between the results of the two approaches, since the value of the critical temperature is reduced even by $70-80 \%$ in the self-consistent t-matrix approach with respect to the mean-field one. 
This difference is expected to show up when comparison with the ongoing experiments will eventually be possible.

\vspace{-0.4cm}
\section{Numerical results \\ for the trapped system} 
\label{sec:trap}
\vspace{-0.2cm}

In this Section, we consider a mass-imbalanced Fermi gas trapped in a harmonic potential $V_\sigma(r)=m_\sigma \omega_\sigma^2 r^2/2$, where the frequencies $\omega_\sigma$ are different for each component $\sigma=(L,H)$. 
It is then convenient to introduce the trap frequency ratio $r_\omega=\omega_L/\omega_H$ between the frequencies of the two components (in particular, $r_\omega=3.60$ for the K-Dy mixture in the experiment of Ref.~\cite{Ravensbergen-Grimm-2020}).

The effects of the harmonic potential are included within a local-density approximation, which partitions the inhomogeneous (trapped) system into locally homogeneous regions with local densities $n_\sigma(\mathbf{r})$ for the two species
(for additional details, cf. Sec.~IV of Ref.~\cite{Pini-2020}).
Without loss of generality, the original anisotropic harmonic potential used in the experiments is then conveniently transformed into an isotropic harmonic potential through a simple rescaling of the spatial coordinates. 
The phase diagrams of the anisotropic and isotropic systems then coincide with each other, provided one replaces each frequency $\omega_\sigma$ of the isotropic system by the corresponding geometric mean 
$(\omega_{\sigma,x} \, \omega_{\sigma,y} \, \omega_{\sigma,z})^{1/3}$ of the frequencies along the three axis of the anisotropic system. 
The only assumption is that the aspect ratio $\lambda_\sigma=\omega_{\sigma,z}/\omega_{\sigma,x}$ of the potentials is the same for the two species (such that the same isotropic mapping can be applied to both species).  
This condition is met by the experiment of Ref.~\cite{Ravensbergen-Grimm-2020}, since both species are trapped by the same laser light.

\begin{figure}[t]
\begin{center}
\includegraphics[width=7.5cm,angle=0]{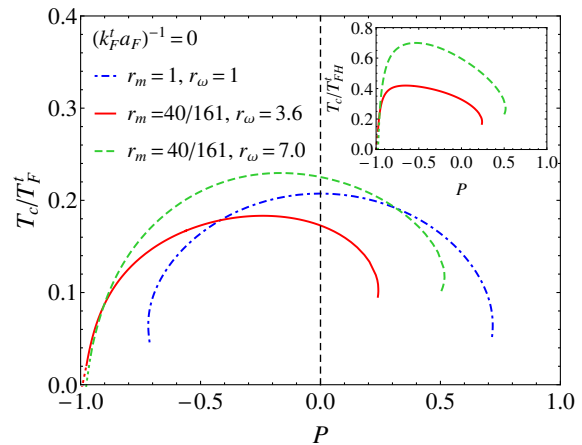}
\caption{(Color online) The critical temperature $T_c$ for the trapped system (in units of the Fermi temperature $T^{t}_F$), obtained at unitarity by the self-consistent $t$-matrix approach, is shown vs the global polarization $P=(N_H-N_L)/(N_H+N_L)$.
                                     Results for a mass-imbalanced K-Dy mixture with trap frequency ratio $r_\omega=3.6$ (full line) and $r_\omega=7.0$ (dashed line) are compared with those of a mass-balanced Fermi gas with equal trap frequencies 
                                     (dashed-dotted line). 
                                     The inset shows the $T_c$ curves for the K-Dy mixture in units of the Fermi temperature $T^{t}_{FH}=\omega_H (6 N_H)^{1/3}/k_B$ of the heavy component.
                                     In main panel and inset, dotted lines are quadratic extrapolations of the K-Dy curves toward $T=0$ for negative $P$.}
\label{Figure-3}
\end{center}
\end{figure} 

Next we define the effective Fermi energy for the trapped system, in the form 
\vspace{-0.1cm}
\begin{equation}
E^{t}_F=\omega_0 (3N)^{1/3} \, ,
\label{Fermi-energy-trapped}
\end{equation}
\noindent
where $\omega_0=(\omega_L \omega_H)^{1/2}$ is the geometric mean of the frequencies for the two species and $N=N_L+N_H$ the total number of atoms. 
Akin to what is done for the homogeneous system, in terms of the Fermi energy (\ref{Fermi-energy-trapped}) for the trapped system we then define the effective Fermi wave vector $k^{t}_F=\sqrt{2 m E^{t}_F}$, the ensuing coupling parameter $(k^{t}_F a_F)^{-1}$, and the Fermi temperature $T^{t}_F=E^{t}_F/k_B$.

In addition, the critical temperature $T_c$ is here defined as the temperature at which the trapped gas becomes superfluid at the trap center. 
Accordingly, we are not searching for the possible occurrence of shell superfluidity away from the trap center \cite{Lin-2006,Wu-Pao-Yip-2006,Wu-Pao-Yip-2007}, 
and limit our investigation to the regions of physical parameters where superfluidity occurs at the trap center only.
We also assume that normal-to-superfluid transition is of second order, as we did in Sec.~\ref{sec:homo} for the homogeneous case.
With these provisions, we are going to determine the dependence of $T_c$ on polarization and frequency ratio.

\vspace{-0.5cm}
\subsection{Dependence of $T_c$ on the polarization $P$}
\label{subsec:Tc_vs_P}
\vspace{-0.2cm}

Figure~\ref{Figure-3}  shows the temperature-polarization phase diagram for a trapped system at unitarity, with the mass ratio $r_m=40/161$ of the K-Dy mixture and the trap frequency ratio $r_\omega=3.6$ 
(like in Ref.~\cite{Ravensbergen-Grimm-2020}) (full line) and $r_\omega=7.0$ (dashed line).
Comparison is also shown with the results for a mass-balanced Fermi gas with equal trap frequencies (dashed-dotted line).
Here $P=(N_H-N_L)/(N_H+N_L)$ is the global population polarization that characterizes the imbalance between the number of atoms of the two species.
In the inset of Fig.~\ref{Figure-3}  the $T_c$ curves for the K-Dy mixture are instead reported in units of the Fermi temperature $T^{t}_{FH}=\omega_H (6 N_H)^{1/3}/k_B$ of the heavy component 
(the conversion factor between the two scales being $T^{t}_{FH}/T^{t}_{F}=(r_\omega)^{-1/2}(1+P)^{1/3}$). 
This is to facilitate future comparisons with the experimental data for the K-Dy mixture, since in the corresponding experiment $T^{t}_{FH}$ is used as the unit of temperature owing to the fact that the heavy component serves for cooling and thermometry purposes 
\cite{Ravensbergen-Grimm-2020}.

\begin{figure}[t]
\begin{center}
\includegraphics[width=7.5cm,angle=0]{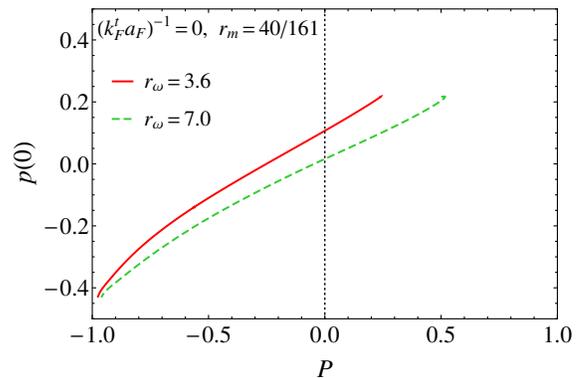}
\caption{(Color online) The local density polarization $p(0)$ at the trap center is shown as a function of the global polarization $P$ for a unitary K-Dy mixture in a harmonic trap at $T_c$, 
                                    with the trap frequency ratio $r_\omega=3.6$ (full line) and $r_\omega=7.0$ (dashed line).}
\label{Figure-4}
\end{center}
\end{figure} 

Similarly to the homogeneous case of Fig.~\ref{Figure-2}, also for the mass-balanced case of Fig.~\ref{Figure-3}  the maximum of $T_c/T^{t}_F$ occurs for equal populations, 
while for the K-Dy mixture the maximum occurs on the $P<0$ side for both frequency ratios $r_\omega$ there considered. 
The value of the polarization $P_\mathrm{max}$ corresponding to this maximum depends on the trap frequency ratio $r_\omega$.
In a simple Thomas-Fermi calculation of the density profiles for the non-interacting mixture at $T=0$, the value of the global polarization $P$ for which the densities at the trap center are equal increases monotonically with $r_\omega$. 
In a related fashion, we also expect the value of $P_\mathrm{max}$ to increase (and eventually change sign) by increasing $r_\omega$, since the maximum value of $T_c$ should correspond to $n_L(r=0) \simeq n_H(r=0)$. 
On the $P>0$ side of the K-Dy phase diagram, Fig.~\ref{Figure-3}  shows that the behavior of the $T_c$ curve is similar to that of the mass-balanced case, with a reentrance at the upper critical polarization $P^{+}$ 
which could signal the presence of phase separation or of exotic phases.
On the $P<0$ side, on the other hand, $T_c$ remains finite down to a lower critical value $P^{-}$ close to $-1$ (specifically, the extrapolated values for the lower critical polarization are
$P^{-}=-0.992$ for $r_\omega=3.6$ and $P^{-}=-0.976$ for $r_\omega=7$) \cite{Footnote-2}.

To better characterize the behavior of $T_c$ vs $P$ for the K-Dy mixture,  Fig.~\ref{Figure-4}  shows the local density polarization  $p(0)=(n_{H}(0)-n_{L}(0))/(n_{H}(0)+n_{L}(0))$ at the trap center 
(with $n_{\sigma}(0)=n_{\sigma}(r=0)$) as a function of the global polarization $P$ for $r_\omega=3.6$ and $r_\omega=7.0$. 
For both values of $r_{\omega}$, the main feature to be noted from Fig.~\ref{Figure-4}  is the wider range of $P$ in contrast to the more compressed range of $p(0)$.
This is because, when the global population imbalance is increased in a trap, the system tends to compensate for this increase and accordingly depletes the minority density profile on the wings to favor pairing at the trap center.

\vspace{-0.5cm}
\subsection{Dependence of $T_c$ on the frequency ratio $r_\omega$}
\label{subsec:r_omega}
\vspace{-0.2cm}

\begin{figure}[t]
\begin{center}
\includegraphics[width=7.5cm,angle=0]{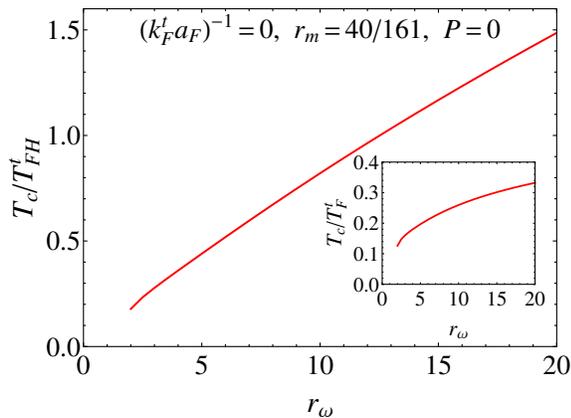}
\caption{(Color online) The critical temperature $T_c$ (in units of the Fermi temperature $T^{t}_{FH}=\omega_H (6 N_H)^{1/3}/k_B$ of the heavy component) for a unitary K-Dy mixture with equal populations ($P=0$) in a harmonic trap is shown
                                     as a function of the trap frequency ratio $r_\omega$. 
                                     In the inset, the same curve is shown with $T_c$ in units of the effective Fermi temperature $T^{t}_{F}=\omega_0 (3 N)^{1/3}/k_B$ for the trapped system.}
\label{Figure-5}
\end{center}
\end{figure} 

Here we analyze dependence of the K-Dy phase diagram on the trap frequency ratio $r_\omega=\omega_L/\omega_H$. 
The range of $r_\omega$ here explored is limited on the small $r_\omega$ side by the occurrence of shell superfluidity. 
We have, in fact, verified that shell superfluidity first appears for $r_\omega \simeq 3.4$ just at the upper critical value $P^+$ where the reentrance occurs in  Fig.~\ref{Figure-3}, and then progressively moves to smaller values of $P$ 
in correspondence to lower values of $r_\omega$, reaching eventually $P=0$ for $r_\omega \simeq 2$. 

Figure~\ref{Figure-5}  shows the critical temperature $T_c$ (in units of the Fermi temperature of the heavy component $T^{t}_{FH}$) as a function of $r_\omega$,
for a trapped K-Dy mixture at unitarity with equal populations ($P=0$). 
The critical temperature is seen to markedly increase for increasing $r_\omega$. 
A physical argument to account for this increase goes as follows. 
\begin{figure}[t]
\begin{center}
\vspace{-0.70cm}
\includegraphics[width=7.5cm,angle=0]{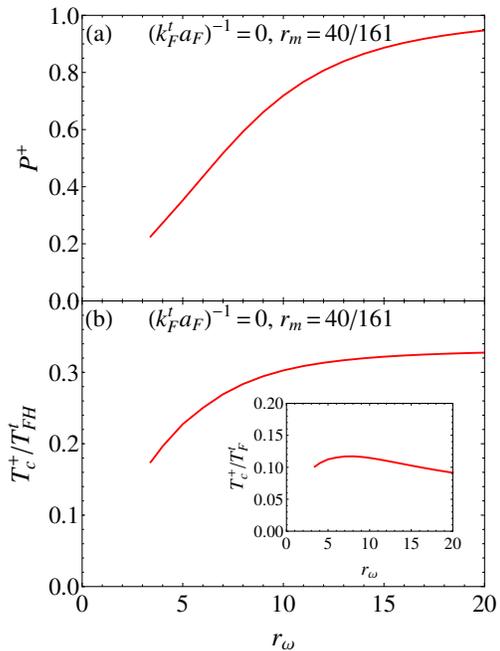}
\caption{(Color online) (a) The global polarization $P^+$ at the reentrance of the $T_c(P)$ curve of the critical temperature for a unitary trapped K-Dy mixture (with $r_m=40/161$) is shown as a function of the trap frequency ratio $r_{\omega}$.
                                     (b) The critical temperature $T^+_c=T_c(P=P^+)$ (in units of $T^{t}_{FH}=\omega_H (6 N_H)^{1/3}/k_B$) at the reentrance of the $T_c(P)$ curve of the critical temperature for a unitary trapped K-Dy mixture is shown 
                                     as a function of the trap frequency ratio $r_{\omega}$. 
                                     In the inset, the same curve is shown with $T_c$ in units of the effective Fermi temperature $T^{t}_{F}=\omega_0 (3 N)^{1/3}/k_B$ for the trapped system.}
\label{Figure-6}
\end{center}
\end{figure} 
\begin{figure}[t]
\begin{center}
\includegraphics[width=7.5cm,angle=0]{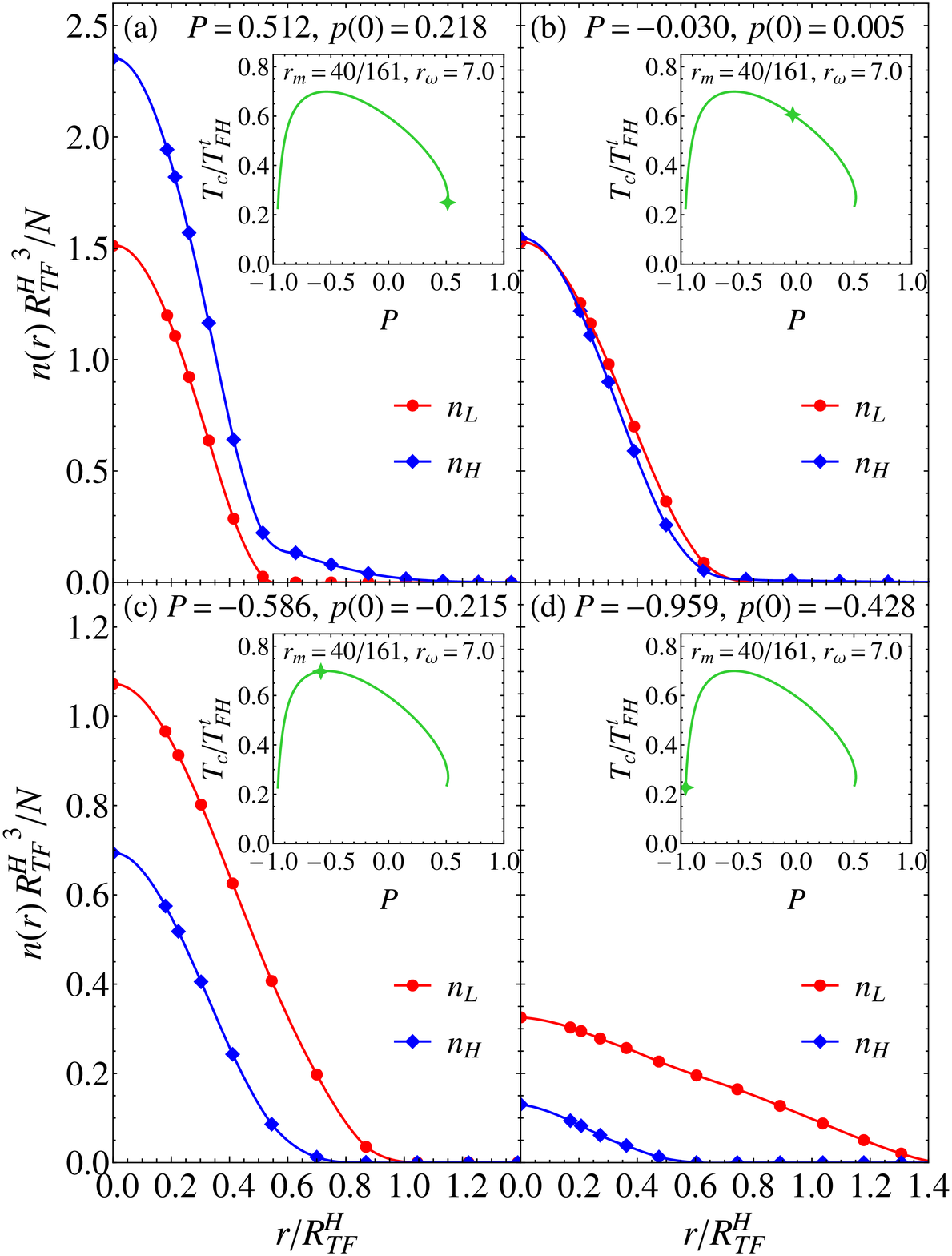}
\caption{(Color online) Density profiles for the light (circles) and heavy (diamonds) components of a unitary K-Dy mixture in a harmonic trap at $T_c$ with a trap frequency ratio $r_\omega=7.0$, for four characteristic values of the global polarization $P$. 
                                    Full lines are interpolations to the calculated points. 
                                    In each panel, the corresponding values of the local polarization $p(0)$ at the trap center are also indicated for convenience.
                                    In the insets, the stars identify the positions along the $T_c(P)$ curves (full lines) of the values of the global polarization used in the main panels for the calculation of the density profiles.}
\label{Figure-7}
\end{center}
\end{figure} 
Suppose that $r_\omega=\omega_L/\omega_H$ gets increased by increasing the trap frequency $\omega_L$ for the light component, while keeping fixed the trap frequency $\omega_H$ for the heavy component in such a way that $T^{t}_{FH}$ remains constant. 
For the lower value $r_\omega=2.0$ of the trap frequency ratio we have considered, our calculation results in a majority of heavy atoms at the trap center corresponding to a positive value of the local density polarization $p(0)$. 
By increasing $r_\omega$, the density of the light atoms at the trap center increases as their trap becomes more confining.
As a consequence, $p(0)$ decreases and eventually approaches zero. 
This situation favors pairing (in accordance with Fig.~\ref{Figure-2}  for the homogeneous case), so that one gets an increase of the critical temperature $T_c/T_F(r=0)$ in units of the local Fermi temperature 
$T_F(r=0)=(3\pi^2 n(0))^{2/3}/(2m k_B)$. 
The total local density $n(0)=n_{L}(0)+n_{H}(0)$ at the trap center increases, too, so that the increase of the critical temperature in units of the trap Fermi temperature of the heavy component $T_c/T^{t}_{FH}=T_c/T_F(r=0)(T_F(r=0)/T^{t}_{FH}$) is going to  
be even more pronounced than that of $T_c/T_F(r=0)$. 
This is because $T_F(r=0)$ is proportional to $n(0)^{2/3}$ while $T^{t}_{FH}$ remains constant. 
We have numerically verified that $T_{c}/T^{t}_{FH}$ continues increasing up to the larger value $r_\omega = 80$ that we have explored (i.e., much beyond the range shown in Fig.~\ref{Figure-5}).

An additional relevant feature of the polarization-temperature phase diagram, to be explored as a function of the trap frequency ratio $r_\omega$, is the reentrance point $(P^+, T^+_c)$ for $P>0$, 
since it provides a rough estimate of the tricritical point at which phase separation begins to appear.
Figures~\ref{Figure-6}(a)  and \ref{Figure-6}(b)  show the dependence on $r_\omega$ of $P^+$ and $T^{+}_c=T_c(P=P^+)$, respectively. 
Upon increasing $r_\omega$, $P^+$ is seen to increase and slowly approach the asymptotic value $P^+ = 1$, while $T^{+}_c/T_{FH}$ appears to saturate at the value of $T^{+}_c/T_{FH} \simeq 0.33$.

\begin{figure}[t]
\begin{center}
\includegraphics[width=7.5cm,angle=0]{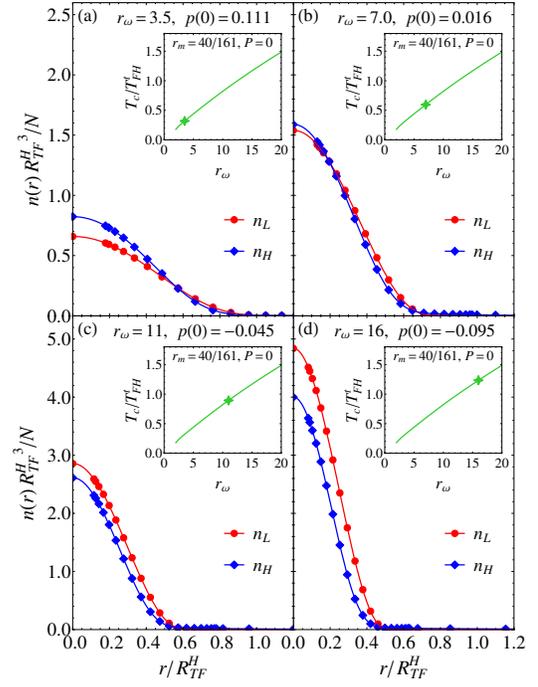}
\caption{(Color online) Density profiles for the light (circles) and heavy (diamonds) components of a unitary K-Dy mixture in a harmonic trap at $T_c$ with equal populations ($P=0$), for four characteristic values of the trap frequency ratio $r_\omega$.
                                     Full lines are interpolations to the calculated points. 
                                     In each panel, the corresponding values of the local polarization $p(0)$ at the trap center are also indicated for convenience.
                                     In the insets, the stars identify the positions along the $T_c(r_\omega)$ curves (full lines) of the values of the trap frequency ratio $r_\omega$ used in the calculation of the density profiles.}
\label{Figure-8}
\end{center}
\end{figure} 

\begin{figure}[t]
\begin{center}
\includegraphics[width=7.5cm,angle=0]{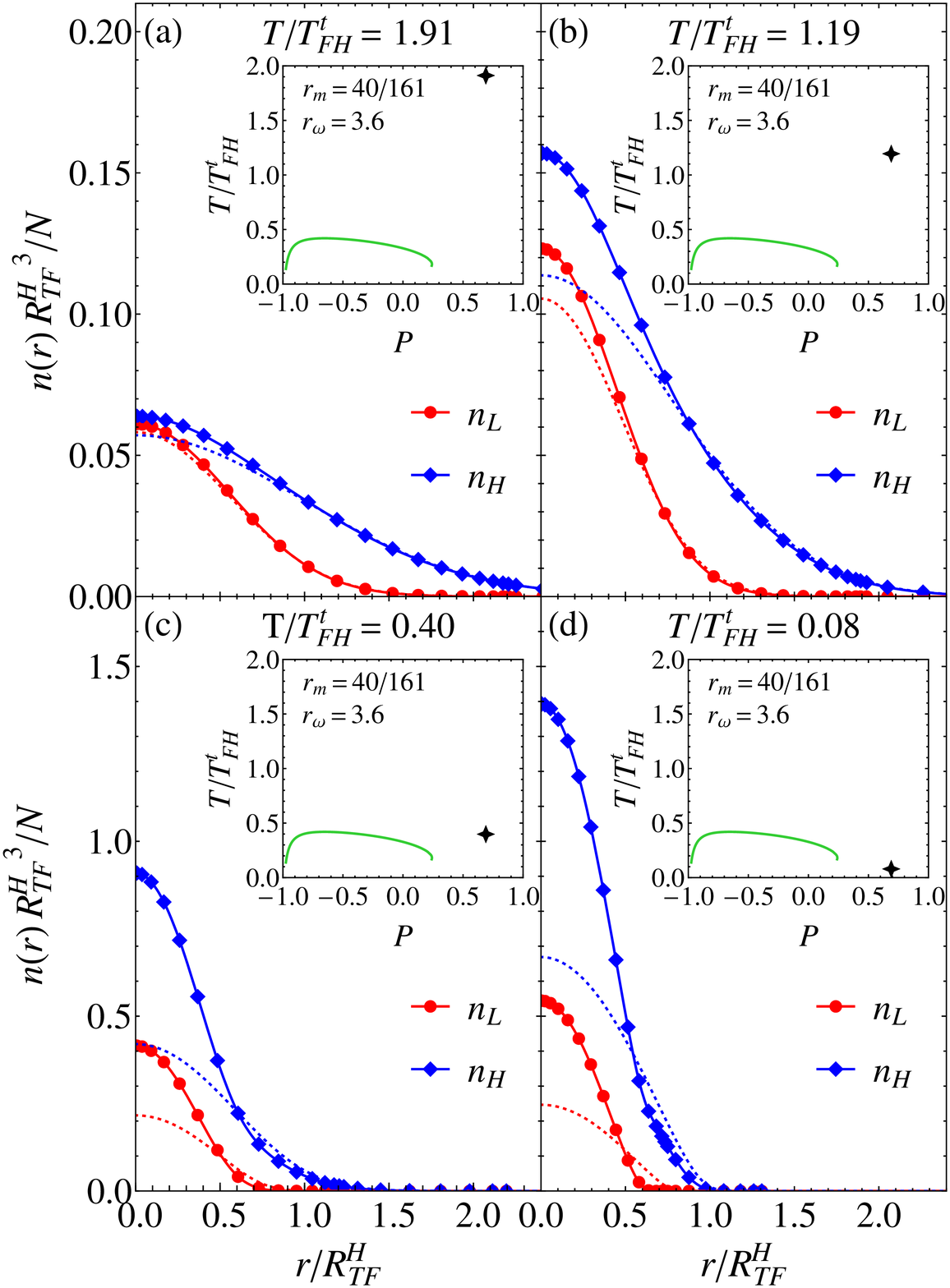}
\caption{(Color online) Density profiles for the light (circles) and heavy (diamonds) components of a unitary K-Dy mixture in a harmonic trap with global polarization $P=0.69$ and trap frequency ratio $r_\omega=3.6$, for four characteristic values of the temperature.
                                    Full lines are interpolations to the calculated points. 
                                    In each case, the corresponding Thomas-Fermi (non-interacting) density profiles are shown for comparison (dotted lines). 
                                    In the insets, the full lines represent the $T_{c}$-vs-$P$ curves and the stars identify the positions in the polarization-temperature phase diagram at which the density profiles of the main panels are calculated.}
\label{Figure-9}
\end{center}
\end{figure} 

\vspace{-0.5cm}
\subsection{Density profiles}
\label{subsec:densityprofiles}
\vspace{-0.4cm}

We now consider representative density profiles calculated both at the critical temperature $T_c$ (as obtained in Secs.~\ref{subsec:Tc_vs_P} and \ref{subsec:r_omega}) and in the normal phase. 
In the following, the radial position $r$ in the trap is conveniently expressed in units of the Thomas-Fermi radius $R_{TF}^{H}$ of the heavy component, as given by the condition $\frac{1}{2}m\omega_H (R_{TF}^{H})^2=E^{t}_{FH}$ 
where $E^{t}_{FH}=\omega_H (6N_H)^{1/3}$ is the Fermi energy for the heavy component.

Figure~\ref{Figure-7}  shows the density profiles calculated at $T_c$ for a unitary K-Dy mixture with $r_\omega=7.0$ (cf. also Fig.~\ref{Figure-3}), for four characteristic values of the polarization $P$. 
For $P>0$ (corresponding to a majority of heavy atoms), the profile of the heavy-atom cloud has a kink where the profile of the light-atom cloud vanishes, with a small tail of non-interacting atoms in excess surviving for larger $r$. 
A similar behavior occurs for $P<0$ (corresponding to a majority of light atoms) for the light-atom cloud, although it appears less evident in this case. 

In addition, Fig.~\ref{Figure-8}  shows the density profiles calculated at $T_c$ for a unitarity K-Dy mixture with global polarization $P=0$ (cf. also Fig.~\ref{Figure-5}) for four characteristic values of the trap frequency ratio $r_\omega$. 
Here, the main effect of increasing $r_\omega$ (which can be obtained by increasing $\omega_L$ and keeping $\omega_H$ fixed) is to shrink both density profiles resulting in an increase the total density at the trap center. 
This is because, as the light-atom cloud gets shrunk by the trap becoming more confining, the heavy-atom cloud gets also shrunk in order to maximize the attractive interaction energy. 
For even larger values of $r_\omega$, however, the heavy-atom cloud will stop following the tight light-atom cloud, because the cost in kinetic energy would overcome the gain in interaction energy. 

Finally, Fig.~\ref{Figure-9}  shows the density profile obtained at various temperatures for a unitary K-Dy mixture with polarization $P=0.69$ and $r_\omega=3.6$ 
(these values of $P$ and  $r_\omega$ correspond to the experimental conditions considered in Fig.~5(a) of Ref.~\cite{Ravensbergen-Grimm-2020}). 
The non-interacting density profiles at the same temperatures are also shown for comparison (dotted lines). 
From this comparison it is evident that at low temperature ($T/T^{t}_{FH}\lesssim1$) the attractive interaction between light and heavy atoms results in a strong contraction of the density profiles with respect to those of the non-interacting case. 

In Ref.~\cite{Ravensbergen-Grimm-2020}, this contraction was proposed as the main mechanism behind the experimentally observed increase of three-body losses of Dy in the K-Dy mixture with respect to the free Dy case. 
To quantify the increase of intra-species three-body recombinations due to the contraction of the density profiles, we follow Ref.~\cite{Ravensbergen-Grimm-2020} (cf. the Supplemental Material therein) and define the factors
\vspace{-0.2cm}
\begin{equation}
\beta_\sigma = \frac{\int d \mathbf{r} \, n^3_\sigma(\mathbf{r})}{\int d\mathbf{r} \, n^3_{\mathrm{TF},\sigma}(\mathbf{r})} \quad (\sigma=L,H)
\label{eq:beta_sigma}
\end{equation}
\noindent
where $n_{\mathrm{TF},\sigma}(\mathbf{r})$ are the corresponding Thomas-Fermi (non-interacting) density profiles.

\begin{figure}[t]
\begin{center}
\includegraphics[width=7.5cm,angle=0]{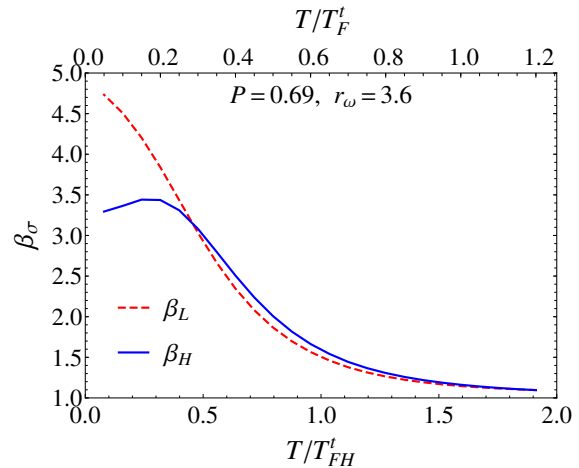}
\caption{(Color online) The three-body decay rate factors $\beta_\sigma$ (with $\sigma=L, H$) for a unitary K-Dy mixture in a harmonic trap with global polarization $P=0.69$ and trap frequency ratio $r_\omega=3.6$ are shown as a function of temperature (for which both
                                     normalizations in terms of $T^{t}_{F}$ and $T^{t}_{FH}$ are used).}
\label{Figure-10}
\end{center}
\end{figure} 

Figure~\ref{Figure-10}  shows the factors $\beta_\sigma$ as a function of temperature for the unitary K-Dy mixture with $P=0.69$ and $r_\omega=3.6$. 
At the lowest temperature ($T/T^{t}_{FH}$=0.08) of our calculation, we obtain the values $\beta_L=4.73$ and $\beta_H=3.29$.
These results imply that in the low-temperature (with respect to $T^{t}_{F H}$) regime of the normal phase the three-body recombination rate can be strongly increased by the contraction of the density profiles. 
[We may mention that the values $\beta^{ph}_L=2.85$ and $\beta^{ph}_H=2.07$ still larger than unity were alternatively obtained by a simple phenomenological model developed at $T=0$ in the Supplemental Material of Ref~\cite{Ravensbergen-Grimm-2020}.] 
At higher temperatures ($T/T^{t}_{FH} \gtrsim 1$), on the other hand, our results imply that the effect of the contraction of the density profiles on the three-body recombinations becomes progressively less relevant, in such a way that $\beta_\sigma$ approaches unity in 
Fig.~\ref{Figure-10}. 
Since a much larger enhancement of Dy losses (of about $4$) was observed in Ref.~\cite{Ravensbergen-Grimm-2020} for $T/T^{t}_{FH}\simeq 1.9$, our calculations do not support the possibility that this large value of $\beta_H$ could result from the contraction of the density profiles.


\vspace{0.2cm}
\section{CONCLUDING REMARKS} 
\label{sec:conclusion}
\vspace{-0.3cm}

We have presented a detailed quantitative analysis of the temperature-polarization phase diagram for the second-order normal-to-superfluid phase transition of a mass-imbalanced Fermi gas at unitarity, for both homogeneous and harmonically trapped systems.
The corresponding phase diagrams have been obtained within a fully self-consistent $t$-matrix approach, which has already proved capable to compare well with quantum Monte Carlo and experimental results for a mass-balanced system at unitarity and is now awaiting comparison with the results of the forthcoming experiments with mass-imbalanced mixtures.
For the homogeneous system we have considered both a $^{6}$Li-$^{53}$Cr and $^{40}$K-$^{161}$Dy mixture, while for the trapped system we have mainly focused on the K-Dy mixture 
since the corresponding experiments have already reached quantum degeneracy and demonstrated tunability.
For the trapped system, we have also investigated the dependence of the phase diagram on the ratio $r_\omega$ of the trap frequencies for the two species, which is an experimentally controllable parameter. 
We have shown that an increase of $r_\omega$ results in an increase of the critical temperature and thus makes the superfluid phase more accessible experimentally.

For the trapped system, we have further presented a number of representative density profiles for the two species, both at $T_c$ and in the normal phase, and we have quantified the effect of the contraction of the density profiles due to the inter-species attraction on the three-body recombination rates with respect to the non-interacting case under the same conditions of the experiment of Ref.~\cite{Ravensbergen-Grimm-2020}. 
Our calculation shows that, although this effect can be important at low temperature (with respect to $T^{t}_{FH}$) in the normal phase, it cannot explain the observed increase of three-body losses at the temperatures considered in Ref.~\cite{Ravensbergen-Grimm-2020}. 

As pointed out in the Introduction, in the present work we did not consider the possible occurrence of a first-order normal-to-superfluid phase transition. 
In practice, this would require a major computational effort, whereby the self-consistent $t$-matrix approach would need to be extended to the superfluid phase in the presence of both population and mass imbalances. 
Nevertheless, a rough estimate of the region of parameters where phase separation (and/or transition to more exotic FFLO phases) is expected to occur, can already be obtained from the position of the reentrance point in the curves of $T_{c}$ that we have presented. 
Specifically, in analogy with what is found at the mean-field level, phase separation is expected to occur for temperatures below and for polarizations close to the reenntrance point in these curves.   
In addition, on the basis of variational arguments, phase separation is expected to enlarge the region of global polarization where superfluidity (coexisting with a normal phase) takes place. 
Even when phase separation occurs, the values we have obtained for the critical polarizations will thus represent a valuable piece of information for the experiments, by providing in this case a lower bound to the actual polarization.

In conclusion, we expect that the results presented in this article can extensively be used as relevant benchmarks for guiding the search of superfluidity in ongoing and future experiments with hetero-nuclear trapped Fermi mixtures.


\vspace{-0.2cm}
\begin{center}
\begin{small}
{\bf ACKNOWLEDGMENTS}
\end{small}
\end{center}
\vspace{-0.2cm}

Partial financial support from the Italian MIUR under Project PRIN2017 (20172H2SC4) is acknowledged.

\appendix   




\begin{thebibliography}{99}

\bibitem{Chin-2010} C.~Chin, R.~Grimm, P.~Julienne, and E.~Tiesinga, \emph{Feshbach resonances in ultracold gases}, Rev. Mod. Phys. {\bf 82}, 1225 (2010). 

\bibitem{Physics-Reports-2018} G. Calvanese Strinati, P. Pieri, G. R\"{o}pke, P. Schuck, and M. Urban, \emph{The BCS-BEC crossover: From ultra-cold Fermi gases to nuclear systems}, Phys. Rep. {\bf 738}, 1 (2018).

\bibitem{Greiner-2003} M.~Greiner, C.~A. Regal, and D.~S. Jin, \emph{Emergence of a molecular Bose-Einstein condensate from a Fermi gas}, Nature {\bf 426}, 537 (2003).

\bibitem{Jochim-2003a} S.~Jochim, M.~Bartenstein, A.~Altmeyer, G.~Hendl, S.~Riedl, C.~Chin, J.~Hecker~Denschlag, and R.~Grimm, \emph{Bose-Einstein condensation of molecules}, Science {\bf 302}, 2101 (2003).
  
\bibitem{Jochim-2003b} S.~Jochim, M.~Bartenstein, A.~Altmeyer, G.~Hendl, C.~Chin, J.~Hecker~Denschlag, and R.~Grimm, \emph{Pure gas of optically trapped molecules created from fermionic atoms}, Phys. Rev. Lett. {\bf 91}, 240402 (2003).

\bibitem{Zwierlein-2003} M.~W. Zwierlein, C.~A. Stan, C.~H. Schunck, S.~M.~F. Raupach, S.~Gupta, Z.~Hadzibabic, and W.~Ketterle, \emph{Observation of Bose-Einstein condensation of molecules}, Phys. Rev. Lett. {\bf 91}, 250401 (2003).

\bibitem{Zwierlein-2006} M.~W. Zwierlein, A.~Schirotzek, C.~H. Schunck, and W.~Ketterle, \emph{Fermionic superfluidity with imbalanced spin populations}, Science {\bf 311}, 492, (2006).

\bibitem{Partridge-2006} G.~B. Partridge, W.~Li, R.~I. Kamar, Y.-a. Liao, and R.~G. Hulet, \emph{Pairing and phase separation in a polarized Fermi gas}, Science {\bf 311}, 503 (2006).

\bibitem{Shin-2008a} Y.-I.~Shin, C.~H. Schunck, A.~Schirotzek, and W.~Ketterle, \emph{Phase diagram of a two-component Fermi gas with resonant interactions}, Nature {\bf 451}, 689 (2008).

\bibitem{Taglieber-2008} M. Taglieber, A.-C. Voigt, T. Aoki, T. W. H\"{a}nsch, and K. Dieckmann, \emph{Quantum degenerate two-species Fermi-Fermi mixture coexisting with a Bose-Einstein condensate}, Phys. Rev. Lett. {\bf 100}, 010401 (2008).

\bibitem{Wille-2008} E. Wille, F. M. Spiegelhalder, G. Kerner, D. Naik, A. Trenkwalder, G. Hendl, F. Schreck, R. Grimm, T. G. Tiecke, J. T. M. Walraven, S.~J.~J.~M.~F. Kokkelmans, E. Tiesinga, and P. S. Julienne, 
                                 \emph{Exploring an ultracold Fermi-Fermi mixture: interspecies Feshbach resonances and scattering properties of $^{6}$Li and $^{40}$K}, Phys. Rev. Lett. {\bf 100}, 053201 (2008).

\bibitem{Voigt-2009} A.-C. Voigt, M. Taglieber, L. Costa, T. Aoki, W. Wieser, T. W. H\"{a}nsch, and K. Dieckmann, \emph{Ultracold heteronuclear Fermi-Fermi molecules}, Phys. Rev. Lett. {\bf 102}, 020405 (2009).

\bibitem{Naik-2011} D. Naik, A. Trenkwalder, C. Kohstall, F. M. Spiegelhalder, M. Zaccanti, G. Hendl, F. Schreck, R. Grimm, T. M. Hanna, and P.S. Julienne, \emph{Feshbach resonances in the $^{6}$Li-$^{40}$K Fermi-Fermi mixture: elastic versus inelastic interactions}, 
                                Eur. Phys. J. D {\bf 65}, 55 (2011).

\bibitem{Ravensbergen-Grimm-2018} C. Ravensbergen, V. Corre, E. Soave, M. Kreyer, E. Kirilov, and R. Grimm, \emph{Production of a degenerate Fermi-Fermi mixture of dysprosium and potassium atoms}, Phys. Rev. A {\bf 98}, 063624 (2018).

\bibitem{Ravensbergen-Grimm-2020} C. Ravensbergen, E. Soave, V. Corre, M. Kreyer, B. Huang, E. Kirilov, and R. Grimm, \emph{Resonantly interacting Fermi-Fermi mixture of $^{161}$Dy and $^{40}$K}, Phys. Rev. Lett. {\bf 124}, 203402 (2020).

\bibitem{Neri-Zaccanti-2020} E. Neri, A. Ciamei, C. Simonelli, I. Goti, M. Inguscio, A. Trenkwalder, and M. Zaccanti, \emph{Realization of a cold mixture of fermionic chromium and lithium atoms}, Phys. Rev. A {\bf 101}, 063602 (2020).

\bibitem{Liu-2003} W. V. Liu and F. Wilczek, \emph{Interior gap superfluidity}, Phys. Rev. Lett. {\bf 90}, 047002 (2003).

\bibitem{Wu-2003} S.-T. Wu and S. Yip, \emph{Superfluidity in the interior-gap states}, Phys. Rev. A {\bf 67}, 053603 (2003).

\bibitem{Bedaque-2003} P. F. Bedaque, H. Caldas, and G. Rupak, \emph{Phase separation in asymmetrical fermion superfluids}, Phys. Rev. Lett. {\bf 91}, 247002 (2003).

\bibitem{Lin-2006} G.-D. Lin, W. Yi, and L.-M. Duan, \emph{Superfluid shells for trapped fermions with mass and population imbalance}, Phys. Rev. A {\bf 74}, 031604(R) (2006).

\bibitem{Wu-Pao-Yip-2006} S.-T. Wu, C.-H. Pao, and S.-K. Yip, \emph{Resonant pairing between fermions with unequal masses}, Phys. Rev. B {\bf 74}, 224504 (2006). 

\bibitem{Wu-Pao-Yip-2007} C.-H. Pao, S.-T. Wu, and S.-K. Yip, \emph{Asymmetric Fermi superfluid with different atomic species in a harmonic trap}, Phys. Rev. A {\bf 76}, 053621 (2007).

\bibitem{Parish-2007} M. M. Parish, F. M. Marchetti, A. Lamacraft, and B. D. Simons, \emph{Polarized Fermi condensates with unequal masses: tuning the tricritical point}, Phys. Rev. Lett. {\bf 98}, 160402 (2007).

\bibitem{Gubbels-2009} K. B. Gubbels, J. E. Baarsma, and H. T. C. Stoof, \emph{Lifshitz point in the phase diagram of resonantly interacting $^{6}$Li-$^{40}$K mixtures}, Phys. Rev. Lett. {\bf 103}, 195301 (2009).

\bibitem{Baarsma-2010} J. E. Baarsma, K. B. Gubbels, and H. T. C. Stoof, \emph{Population and mass imbalance in atomic Fermi gases}, Phys. Rev. A {\bf 82}, 013624 (2010).

\bibitem{Wang-2017} J. Wang, Y. Che, L. Zhang, and Q. Chen, \emph{Enhancement effect of mass imbalance on Fulde-Ferrell-Larkin-Ovchinnikov type of pairing in Fermi-Fermi mixtures of ultracold quantum gases}, Sci. Rep. {\bf 7}, 39783 (2017).

\bibitem{Casalbuoni-Nardulli-2004} R. Casalbuoni and G. Nardulli, \emph{Inhomogeneous superconductivity in condensed matter and QCD}, Rev. Mod. Phys. {\bf 76}, 263 (2004).

\bibitem{Gezerlis-2009} A. Gezerlis, S. Gandolfi, K. E. Schmidt, and J. Carlson, \emph{Heavy-light fermion mixtures at unitarity}, Phys. Rev. Lett. {\bf 103}, 060403 (2009).


\bibitem{Zdybel-2018} P. Zdybel and P. Jakubczyk, \emph{Effective potential and quantum criticality for imbalanced Fermi mixtures}, J. Phys. Condens. Matter {\bf 30}, 305604 (2018).

\bibitem{Zdybel-2019} P. Zdybel and P. Jakubczyk, \emph{Damping of the Anderson-Bogolyubov mode in Fermi mixtures by spin and mass imbalance}, Phys. Rev. A {\bf 100}, 053622 (2019).

\bibitem{Zdybel-2020} P. Zdybel and P. Jakubczyk, \emph{Quantum Lifshitz points and fluctuation-induced first-order phase transitions in imbalanced Fermi mixtures}, Phys. Rev. Res. {\bf 2}, 033486 (2020).


\bibitem{Hanai-JLTP-2013} R. Hanai, T. Kashimura, R. Watanabe, D. Inotani, and Y. Ohashi, \emph{Superfluid phase transition and strong-coupling effects in an ultracold Fermi gas with mass imbalance}, J. Low Temp. Phys. {\bf 171}, 389 (2013).

\bibitem{Hanai-PRA-2013} R. Hanai, T. Kashimura, R. Watanabe, D. Inotani, and Y. Ohashi, \emph{Excitation properties and effects of mass imbalance in the BCS-BEC crossover regime of an ultracold Fermi gas}, Phys. Rev. A {\bf 88}, 053621 (2013).

\bibitem{Hanai-JLTP-2014} R. Hanai and Y. Ohashi, \emph{Self-consistent t-Matrix approach to an interacting ultracold Fermi gas with mass imbalance}, J. Low Temp. Phys. {\bf 175}, 272 (2014). 

\bibitem{Hanai-PRA-2014} R. Hanai and Y. Ohashi, \emph{Heteropairing and component-dependent pseudogap phenomena in an ultracold Fermi gas with different species with different masses}, Phys. Rev. A {\bf 90}, 043622 (2014).

\bibitem{Haussmann-1993} R. Haussmann, \emph{Crossover from BCS superconductivity to Bose-Einstein condensation: A self-consistent theory}, Z. Phys. B {\bf 91}, 291 (1993).

\bibitem{Haussmann-1994} R. Haussmann, \emph{Properties of a Fermi liquid at the superfluid transition in the crossover region between BCS superconductivity and Bose-Einstein condensation}, Phys. Rev. B {\bf 49}, 12975 (1994).

\bibitem{Haussmann-2008} R.~Haussmann and W.~Zwerger, \emph{Thermodynamics of a trapped unitary Fermi gas}, Phys. Rev. A {\bf 78}, 063602 (2008).

\bibitem{PPS-2019} M. Pini, P. Pieri, and G. Calvanese Strinati, \emph{Fermi gas throughout the BCS-BEC crossover: Comparative study of $t$-matrix approaches with various degrees of self-consistency}, Phys. Rev. B {\bf 99}, 094502 (2019).

\bibitem{Ku-Zwierlein-2012} M.~J.~H. Ku, A.~T. Sommer, L.~W. Cheuk, and M.~W. Zwierlein, \emph{Revealing the superfluid lambda transition in the universal thermodynamics of a unitary Fermi gas}, Science {\bf 335}, 563 (2012).

\bibitem{Jensen-2020} S. Jensen, C. N. Gilbreth, and Y. Alhassid, \emph{Contact in the unitary Fermi gas across the superfluid phase transition}, Phys. Rev. Lett. {\bf 125}, 043402 (2020).

\bibitem{Pini-2020} M. Pini, P. Pieri,  M. J\"ager, J. Hecker Denschlag, and G. Calvanese Strinati, \emph{Pair correlations in the normal phase of an attractive Fermi gas}, New J. Phys. {\bf 22}, 083008 (2020).

\bibitem{Thouless-1960} D. J. Thouless, \emph{Perturbation theory in statistical mechanics and the theory of superconductivity}, Ann. Phys. {\bf 10}, 553 (1960).

\bibitem{Efimov-1970} V. Efimov, \emph{Energy levels arising from resonant two-body forces in a three-body system}, Phys. Lett. B {\bf 33}, 563 (1970).

\bibitem{Efimov-1973} V. Efimov, \emph{Energy levels of three resonantly interacting particles}, Nucl. Phys. A {\bf 210}, 157 (1973).

\bibitem{Petrov-2003} D. S. Petrov, \emph{Three-body problem in Fermi gases with short-range interparticle interaction}, Phys. Rev. A {\bf 67}, 010703(R) (2003).

\bibitem{Petrov-2005} D. S. Petrov, C. Salomon, and G. V. Shlyapnikov, \emph{Diatomic molecules in ultracold Fermi gases-Novel composite bosons},  J. Phys. B: At. Mol. Opt. Phys. {\bf 38}, S645 (2005). 

\bibitem{Frank-2018} B.~Frank, J.~Lang, and W.~Zwerger, \emph{Universal phase diagram and scaling functions of imbalanced Fermi gases}, J. Exp. Th. Phys. {\bf 127}, 812 (2018).

\bibitem{Footnote-1} Our convention here that $p$ is positive for $n_H>n_L$ is opposite to that of Refs.~\cite{Gubbels-2009,Baarsma-2010}.

\bibitem{Sarma-1963} G. Sarma, \emph{On the influence of a uniform exchange field acting on the spins of the conduction electrons in a superconductor}, J. Phys. Chem. Solids {\bf 24}, 1029 (1963).

\bibitem{Footnote-2} That $P^{-} \rightarrow -1$ when $r_{m} = 0.25$ was also obtained in Ref.~\cite{Braun-2014}, in terms of a phenomenological approach that combines a variational procedure with quantum Monte Carlo data.

\bibitem{Braun-2014} J.~Braun, J.~E.~Drut, T.~Jahn, M.~Pospiech, and D.~Roscher, \emph{Phases of spin- and mass-imbalanced ultracold Fermi gases in harmonic traps}, Phys. Rev. A {\bf 89}, 053613 (2014).

\end{thebibliography}
\end{document}